\documentclass[11pt,reqno]{amsart}
\usepackage{geometry}
\usepackage{graphicx}
\usepackage{amsmath}
\usepackage{amssymb}
\usepackage{amsaddr}
\usepackage{epstopdf}
\usepackage{color}
\usepackage{amsthm}
\usepackage{diagbox}
\usepackage{multirow}
\usepackage[section]{algorithm}
\usepackage{float}
\usepackage[round, sort]{natbib}
\usepackage{enumitem}
\setlist[itemize]{align=parleft,left=1pt..2.5em}

\makeatletter
\newenvironment{breakablealgorithm}
  {
   \begin{center}
     \refstepcounter{algorithm}
     \hrule height.8pt depth0pt \kern2pt
     \renewcommand{\caption}[2][\relax]{
       {\raggedright\textbf{\fname@algorithm~\thealgorithm} ##2\par}%
       \ifx\relax##1\relax 
         \addcontentsline{loa}{algorithm}{\protect\numberline{\thealgorithm}##2}%
       \else 
         \addcontentsline{loa}{algorithm}{\protect\numberline{\thealgorithm}##1}%
       \fi
       \kern2pt\hrule\kern2pt
     }
  }{
     \kern2pt\hrule\relax
   \end{center}
  }
\makeatother

\title[Similarity of competing risks models with constant intensities]{Similarity of competing risks models with constant intensities in an application to clinical healthcare pathways involving prostate cancer surgery}
\author{Nadine Binder}
\address{Institute for General Practice/Primary Care, Medical Center and Faculty of Medicine, University of Freiburg, Germany}

\author{Kathrin Möllenhoff}
\address{Mathematical Institute, Heinrich-Heine University, Düsseldorf, Germany; correspondence to kathrin.moellenhoff@hhu.de}

\author{August Sigle}
\address{Department of Urology, Faculty of Medicine, Medical Center - University of Freiburg, Freiburg, Germany}

\author{Holger Dette}
\address{Department of Mathematics, Ruhr-Universität Bochum, Germany}

\date{\today}                                    

\begin{document}

\begin{abstract}
The recent availability of routine medical data, especially in a university-clinical context, may enable the discovery of typical healthcare pathways, i.e., typical temporal sequences of clinical interventions or hospital readmissions. However, such pathways are heterogeneous in a large provider such as a university hospital, and it is important to identify similar care pathways that can still be considered typical pathways. We understand the pathway as a temporal process with possible transitions from a single initial treatment state to hospital readmission of different types, which constitutes a competing risk setting. In this paper, we propose a multi-state model-based approach to uncover pathway similarity between two groups of individuals. We describe a new bootstrap procedure for testing the similarity of transition intensities from two competing risk models with constant transition intensities. In a large simulation study, we investigate the performance of our similarity approach with respect to different sample sizes and different similarity thresholds. The studies are motivated by an application from urological clinical routine and we show how the results can be transferred to the application example.
\end{abstract}

\maketitle

\section{Introduction}\label{intro}

In the context of evidence-based medicine and guidelines, there is still a high degree of unwarranted differences in individual disease-specific healthcare pathways. A healthcare pathway can be broadly seen as the route that a patient follows from the first contact with a medical doctor, e.g., the general practitioner, through referral to specialists or hospitals to the completion of treatment for any specific disease. It is a timeline in which all treatment-related events can be entered, including diagnoses, treatments, and further consultations or hospital re-admissions. 
The novel availability of medical routine data, especially in the university-clinical context, not only makes it possible to show differences in treatment. Rather, it may also allow to uncover typical clinical healthcare pathways, i.e., typical temporal sequences of clinical interventions or readmissions into the clinic, and to make them available to other clinicians in context. This could enable to improve general standards of clinical care and thus overall health outcomes. 
Still, pathways of patients in a large provider as a university hospital are heterogeneous as many diagnostic and treatment options exist and patients are partly readmitted to the hospital after discharge for different reasons. A {\it similar} healthcare pathway could still be considered a {\it typical} healthcare pathway. For this purpose, key questions would be how to measure such similarity and how to decide whether two different paths are still similar and when they would be considered different. To date, very few methodological works on the measurement of healthcare similarity can be found and these are predominantly informatics-based. For instance, assuming that healthcare pathways depend on factors such as choices made by the treating physician, \citet{huang_similarity_2014} suggest a fully unsupervised algorithmic approach based on a probabilistic graphical model representing a mixture of treatment behaviors by latent features.

From a clinical and also patient-centered perspective, it is essential to keep the care pathway as short as possible and prevent complications or disease-related hospital readmissions. 
In this paper, motivated by an application from urologic clinical practice, we would therefore like to focus on objective and universally-recorded clinical event measures including main events `hospital treatment' and `hospital readmission'. We understand the pathway as a temporal process with possible transitions from a single initial treatment state to hospital readmission of different type. We consider the time-to-first hospital readmission, whichever comes first, which constitutes a competing risks setting \citep{andersen_competing_2002}. Specifically, we aim to judge similarity of such pathways for samples of two different populations: group (i) patients {\it receiving} specific inhouse diagnostics before hospital treatment, and group (ii) patients {\it not receiving} specific inhouse diagnostics before hospital treatment. 
Our interest in the similarity of these pathways has the following reasons: While a certain disease requires specific treatment that is often only offered in specialized clinics, diagnostical tools are often more diverse and partly offered in outpatient facilities. Therefore, treatment data including diagnostics performed are often not readily available from the non-clinical sector (at least in Germany) and can not yet or only insufficiently be used for the investigation of healthcare pathway similarities. From such a path perspective, one may ask whether it makes a difference in terms of hospital readmission whether a particular diagnostic procedure was performed in the clinic or not. From the perspective of the clinical practitioner, it may be plausible to assume that the probability of hospital readmission differs only by treatment, not by different pre-treatment diagnostics. If we could statistically show a similarity of the pathways of both groups, the latter assumption would be confirmed and we may attribute similar pathways to typical pathways.

In this paper, we propose a multi-state model-based approach to reveal such path similarities of two groups of individuals. Multi-state models based on counting processes for event history data have been successfully applied to analyze progression of a disease \citep{andersen_statistical_1993, andersen_multi-state_2002,manzini_advantages_2018}. In the context of care pathways or similarity, however, they have been used only rarely so far and for other purposes. \citet{gasperoni_multi-state_2017} investigated multi-state models for evaluating the impact of risk factors on heart failure care paths involving multiple hospital admissions, admissions to home care or intermediate care units or death. \citet{gasperoni_evaluating_2020} considered potential similarities and differences among healthcare providers on the clinical path of heart failure patients. 

Our approach differs from this work and we aim for testing the similarity of the transition intensities from two independent competing risks Markov models with constant intensities. Then, the problem is methodologically related to the meanwhile classical problem of {\it bioequivalence}, which aims at demonstrating the similarity between two pharmacokinetic profiles by considering the area under the curve or the maximum concentrations of the two curves (see the monographs \citet{chowliu1992,wellek2010testing} among many others). However, none of these methods for establishing bioequivalence can be transferred to the comparison of transition intensities as they are usually developed under the assumption of normally distributed (independent) data. Further, although the asymptotic distribution could be derived for this case as well, an approach based on asymptotics would not yield satisfying power for small sample sizes or data with only few events.  In fact in the following  we will develop new bootstrap  methodology to address this problem.

The paper is structured as follows: Section \ref{appl} describes the clinical healthcare pathways in the application example involving prostate cancer surgery. Section \ref{crm} introduces the competing risks notation for samples of two different populations. In Section \ref{simapp} we describe a novel bootstrap procedure for testing similarity of transition intensities from two competing risks models. In a large simulation study created on the basis of the numbers and estimates from the application example we investigate the performance of our similarity approach with respect to different sample sizes and different similarity thresholds (Section \ref{sim}). In Section \ref{appl2} we briefly discuss how the results from the simulation study translate to the application example. We close the paper with a discussion in Section \ref{disc}.

\section{Clinical healthcare pathways involving prostate cancer surgery} \label{appl}

The application example that drove our methodological development comes from the clinical practice of the Department of Urology at the Medical Center-University of Freiburg. The clinic covers the entire spectrum of urological diagnostics and therapy according to the current state of the art. As data basis, we use the German reimbursement claims dataset for inpatient healthcare, which was systematically integrated into a central database at the Medical Center-University of Freiburg as part of the German Medical Informatics Initiative.   
For each inpatient case, the admission and discharge diagnoses (main and secondary diagnoses) are coded in the form of ICD10 (10th revision of the International Statistical Classification of Diseases and Related Health Problems) codes; in addition, all applied and billing-relevant diagnostic and therapeutic procedures are coded together with a time stamp in the form of OPS (``Operationen- und Prozedurenschl{\"u}ssel'') codes. 

\subsection{Hospital readmission after surgery with and without prior fusion biopsy}

One of the most frequent reasons for inpatient admission at the Department of Urology is prostate cancer. One possible treatment option is the open or robotic-assisted surgery with the resection of the prostatic gland along with the vesicular glands, also referred to as radical prostatectomy \citep{mottet_eau-estro-siog_2017}. From our reimbursement claims database, we retrospectively identified all patients with prostate cancer who underwent {\it open radical prostatectomy} (ORP) at the Medical Center - University of Freiburg between 01 January 2015 and 01 February 2021. This includes all cases with OPS code 5-604 (radical prostatovesiculectomy) irrespective of the concrete surgical procedure -- but without the additional OPS code 5-987 for robotic assistance -- and resulted in a total of n=695 patients. 

\begin{figure}[t]
\begin{center}
\includegraphics[width=\textwidth]{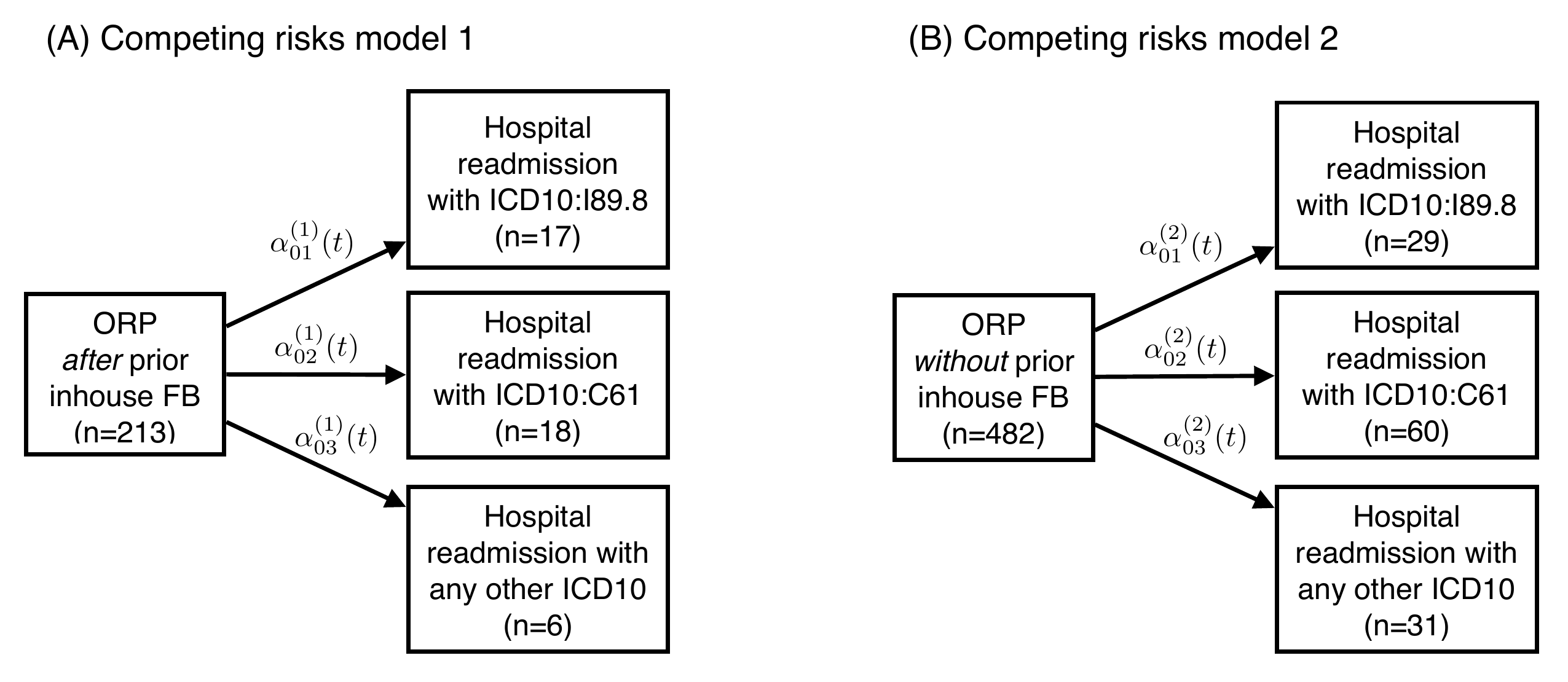}
\caption{Competing risks multi-state models illustrating healthcare pathways for two populations: (A) patients {\it receiving} inhouse fusion biopsy prior to open radical prostatectomy and (B) patients {\it not receiving} inhouse fusion biopsy prior to open radical prostatectomy. The arrows indicate the transitions between the states that are investigated. The $\alpha_{0j}^{(\ell)}$, $ j=1,2,3,\ \ell=1,2$ mark the transition intensities.}
\label{uromsm}
\end{center}
\end{figure}

Prior to surgical intervention, diagnostics are performed in a variety of ways both in the clinic or in an out-of-hospital setting. The current diagnostic standard is a multiparametric magnetic resonance imaging-based pathway with targeted fusion biopsy (FB; OPS code 1-465). However, only a part of the patients receives their biopsies at the Department of Urology, which often depends on the practice of the referring urologists in private practices. In our data n=213 (31\%) patients received FB diagnostic prior to ORP, while a larger part of the patients, n=482 (69\%), did not receive FB diagnostic at the Department of Urology prior to ORP. We did not place a time restriction on when exactly the FB diagnostic took place prior to ORP. Therefore, we distinguish the two populations based on the pre-surgery FB diagnostic obtained and are interested in their further healthcare paths regarding hospital readmission by means of the two independent competing risks models as illustrated in Figure~\ref{uromsm}. 

Radical prostatectomy carries a risk of postoperative complications. One of the more common complications is lymphocele, which typically develops within a couple of weeks after surgery and can be treated at the Department of Urology. Patients may, however, be also readmitted to the clinic after radical prostatectomy for other reasons related to the initial surgery. A typical time window for surgery-related hospital readmission is 90 days after surgery. Therefore, competing outcomes of interest are different reasons for hospital readmission within 90-days. In the data, we identified the most frequent readmission diagnoses defined by the ICD10 main diagnosis codes ``C61: Malignant neoplasm of prostate" (model 1: n=18, 8\%; model 2: n=60, 12\%), ``I89.8: Other specified noninfective disorders of lymphatic vessels and lymph nodes" (model 1: n=17, 8\%; model 2: n=29, 6\%), and combined all other observed diagnoses into one state ``any other diagnosis" (model 1: n=6, 3\%; model 2: n=31, 6\%). While I89.8 is typically coded for a complication after radical prostatectomy requiring specific treatment, a readmission with a C61 diagnosis may mask diagnostic procedures after surgery only or specific follow-up treatment. For all patients in our data set we have at least 90 days clinical follow-up information available, so censoring is only administrative at 90 days after ORP. All cases had to be complete, that is, a discharge date for the initial stay with ORP as well as a potential readmission stay had to be present at the time of data retrieval. 

We note here that the terms $\alpha_{0j}^{(l)}(t)$ in Figure~\ref{uromsm} describe the transition intensities to move from the initial state (ORP) into any of the competing states (hospital readmission with ICD10:I89.8, ICD10:C61, or any other ICD10) and are central in this work. They are formally defined in equation~\eqref{intensities} in Section~\ref{msm}. Assuming the transition intensities to be constant over time, one may estimate them separately by dividing the sum of type-j-events through the sum of person-time at risk in the initial state (see equation~\eqref{mle} in Section~\ref{msm} for a formal definition). In our data, this yields the following estimates: 
\begin{align}\label{estimates}
    \hat\alpha_{01}^{(1)} = 0.001,\quad \hat\alpha_{02}^{(1)} &= 0.0011,\quad \hat\alpha_{03}^{(1)} = 0.0004 \nonumber\\
    \hat\alpha_{01}^{(2)} = 0.0008,\quad \hat\alpha_{02}^{(2)} &= 0.0017,\quad \hat\alpha_{03}^{(2)} = 0.0009 
\end{align}

The estimated constant intensities should be interpreted in the context of the scale in which the time was measured. In our case, the time was measured in days. Since the observation period for all patients was identical (90 days) and overall only few events were observed, the magnitude of the intensities can be appropriately converted into an approximate number of expected events using the formula: transition intensity estimate times observation period (in days) times sample size of the population under consideration (compare eq. \eqref{mle} for the precise definition of the estimate). For example, for transition intensity estimate $\hat\alpha_{01}^{(1)}$ this means $0.001 * 90 \mbox{~days} * 213 \mbox{~patients} \approx 19 \mbox{~events}$.

As the readmission intensities are overall low, from a pathway analytic perspective the question is whether they are sufficiently similar for patients {\it with} prior in-house FB diagnostics versus {\it without} prior in-house FB diagnostics w.r.t. the specific transition such that the two populations can be combined, e.g., for a common analysis on hospital readmission due to complications.

\section{Similarity of competing risks processes for two populations} \label{msm}
\subsection{Competing risk models} \label{crm}

To model the event histories as competing risks for samples of two different populations, we use two independent Markov processes 
 \begin{equation}
     \label{model}
(X^{(\ell )}(t))_{t \geq 0} \quad\quad (\ell  =1,2)
 \end{equation}
 with state spaces $\{0, 1, \ldots, k\}$ following \citet{andersen_statistical_1993}. The processes have possible transitions from state $0$
 to state $j \in \{1, \ldots, k\}$ with transition probabilities 

\begin{equation}
\mathbb{P}^{(\ell  )}_{0j}(0,t) = \mathbb{P}(X^{(\ell  )}(t) = j | X^{(\ell  )}(0) = 0).
\end{equation}

Every individual starts in state $0$ at time $0$, i.e. $P(X(0)=0)=1$. The time-to-first-event is defined as stopping time $T = \inf\{t > 0 \mid X(t) \neq 0\}$ and the type of the first event is $X(T) \in {1,\ldots, k}$. Let

\begin{equation} \label{intensities}
\alpha_{0j}^{(\ell  )}(t) = \lim_{\Delta t \rightarrow 0} \frac{\mathbb{P}^{(\ell  )}_{0j}(t,t + \Delta t)}{\Delta t} \quad (j=1, \ldots , k)
\end{equation}
denote the cause-specific transition intensity from state $0$ to state $j \in \{1, . . . , k\}$ for the  $\ell$th model,
$\ell   \in \{1, 2\}$. The transition intensities completely determine the stochastic behavior of the process. 
In our application example, the two competing risk models with the initial state ORP and three competing risks each are shown in Figure~\ref{uromsm}, in which the transition intensities are assigned to the transition arrows. 

\subsection{Similarity of competing risk models} \label{simapp}

We are interested in the similarity between the transition intensities in the two models. 
 In other words, we want to test the hypotheses
\begin{equation}\label{h2}
  H_0 : \mbox{ there exists an index } j \in \{ 1,\ldots,k \} \mbox{ such that } \| \alpha^{(1)}_{0j} - \alpha^{(2)}_{0j} \|_\infty \geq \Delta_j
\end{equation}
versus
\begin{equation}\label{h3}
  H_1 : \mbox{ for all } j \in \{ 1,\ldots,k \} \qquad  \| \alpha^{(1)}_{0j} - \alpha^{(2)}_{0j} \|_\infty < \Delta_j.
\end{equation}
Here $\| f-g\|_\infty = \sup_{t \in \mathcal{T}} \mid f(t) - g(t) \mid$ denotes the maximal deviation
between the functions $f$ and $g$ and $\Delta_1 , \ldots , \Delta_k$ are pre-specified thresholds, defining for each pair of  transition intensities  the maximum deviation $\Delta_j$ under which $\alpha^{(1)}_{0j} $ and $ \alpha^{(2)}_{0j}$ are considered as similar.

 In order to make the method easily understandable and to be able to provide closed form solutions for the estimates (for a discussion on that, see for example \citet{von2017}) we will assume constant transition intensities  throughout this paper. This assumption is frequently made in the literature (see for instance \citet{fay2003,choudhury2002} among many others). For the same reason
 we restrict ourselves to the case of no censoring (see Section \ref{censoring} for a brief discussion of the right-censored case).


In the following, we describe a novel bootstrap procedure 
for testing the hypotheses  \eqref{h2} and \eqref{h3}
for  competing risk models with constant transition intensities, which is motivated by the methodology developed in
\citet{detmolvolbre2015} for comparing regression curves. 
To be precise, assume that two independent samples $X^{(1)}_1, \ldots, X^{(1)}_{n_1} $ and  $X^{(2)}_1, \ldots, X^{(2)}_{n_2}$
from  Markov processes \eqref{model}
are observed over the interval $\mathcal{T}=[0,\tau]$, containing the state and transition time of an individuals $i$. We
define  
$$
N^{(\ell),i}_{0j} (\tau) 
=
\begin{cases}
1 & \mbox{ if there is a transitions from  } 0 \mbox { to  } j \mbox { in } [0,\tau]  \\
0 &   \mbox{ else }
\end{cases}
$$
as the indicator that  a state transition of the individual $i$ from $0$ to  $j$ has occurred in the time interval $[0,\tau]$ 
(note  that $N^{(\ell),i}_{0j}(\tau) $ 
is either $0$ or $1$). 
We also denote by
$
0 < T^{(\ell),i}_{0j} \leq \tau
$
the corresponding transition time (if $N^{(\ell),i}_{0j}(\tau) =0$  the transition time is undefined).
Further we introduce the notation 
$$Y^{(\ell),i}_0(t)=I\{ X^{(\ell)}_i(t-) =0 \},$$
which indicates whether at time $t$ the $i$th individual of the $\ell$th group is at risk or not. 
Under the assumption of constant transition intensities it then  follows from \citet{andersen_multi-state_2002} that the corresponding likelihood function  in the $\ell$th model 
is given by
\begin{eqnarray}\label{h9}
\mathcal{L}_\ell (\alpha^{(\ell)}) & = &
   \prod_{j=1}^{k}  \prod_{i=1}^{n_\ell}  \big ( \alpha^{(\ell)}_{0j} \big )^{
    N^{(\ell),i}_{0j} (\tau) }
   \exp \big( -  \alpha^{(\ell)}_{0j} \int_{0}^{\tau}Y^{(\ell),i}_0(t)dt \big ) \\
     & = &   \prod_{j=1}^{k}   \big ( \alpha^{(\ell)}_{0j} \big )^{
   N^{(\ell)}_{0j} (\tau) }
   \exp \big( -  \alpha^{(\ell)}_{0j}
    S^{(\ell)}_0 \big )~, 
   \nonumber
\end{eqnarray}
where 
\begin{equation}\label{h5}
  N^{(\ell)}_{0j} (\tau) = \sum_{i=1}^{n_\ell} N^{(\ell),i}_{0j}(\tau)
\end{equation}
is  the number of transitions from state $0$ to state $j$ 
in the $\ell$th group, 
 $$S^{(\ell)}_0= \sum_{i=1}^{n_\ell}\int_{0}^{\tau}Y^{(\ell),i}_0(t)dt$$
 is the total observation time of all individuals in the $\ell$th group,
$\alpha^{(\ell)} = ( \alpha^{(\ell)}_{01} , \ldots ,\alpha^{(\ell)}_{0k})^\top $ is the vector 
of transition intensities in model $\ell=1,2$ and $I\{A\} $ denotes the indicator of the event $A$.
The logarithm of \eqref{h9} is given by
\begin{equation}\label{loglikelihood}
\log \mathcal{L}_\ell (\alpha^{(\ell)})=\sum_{j=1}^k \log(\alpha^{(\ell)}_{0j}) N^{(\ell)}_{0j} (\tau)-\alpha^{(\ell)}_{0j}S^{(\ell)}_0.
\end{equation}
Taking the partial derivatives and equating to zero yields the maximum likelihood estimates (MLE)
\begin{equation}\label{mle}
\hat  \alpha^{(\ell)}_{0j}= \tfrac{N^{(\ell)}_{0j} (\tau)}{S^{(\ell)}_0}\ (j=1,\ldots k,\ \ell=1,2).
\end{equation}
Via $S^{(\ell)}_0$ in \eqref{mle} the intensity estimate depends on the time scale, as already pointed out at the end of Section \ref{appl}.
We now want to address the question of similarity as stated in the hypotheses \eqref{h2} and \eqref{h3}. Due to the assumption of constant transition intensities the maximum deviation simplifies  to 
$$\| \alpha^{(1)}_{0j} - \alpha^{(2)}_{0j} \|_\infty=| \alpha^{(1)}_{0j} - \alpha^{(2)}_{0j}|
$$ that is we consider the absolute difference between these intensities for all states $j=1,\ldots,k$.
In order to reject the null hypothesis in \eqref{h2} the differences between transition intensities have to be smaller than the pre-specified margins $\Delta_j$ for all states. Hence the test problem can be assessed by simultaneously testing the individual hypotheses
\begin{equation}\label{h2ind}
  H^j_0 : | \alpha^{(1)}_{0j} - \alpha^{(2)}_{0j} | \geq \Delta_j
\end{equation}
versus
\begin{equation}\label{h3ind}
  H^j_1 : | \alpha^{(1)}_{0j} - \alpha^{(2)}_{0j} | < \Delta_j
\end{equation}
for all $j=1,\ldots k$.
According to the intersection union principle \citep{berger1982} the global null hypothesis in \eqref{h2} can be rejected at a significance level of $\alpha$ if the individual null hypotheses in \eqref{h2ind} are rejected at a significance level of $\alpha$ for all $j=1,\ldots k$. This means in particular that there is no adjustment of the level necessary.
The following algorithm summarizes how these individual tests are performed.\\ 

\begin{breakablealgorithm} 
\caption{Similarity of transition intensities via constrained parametric bootstrap}
\label{alg1} 
\begin{itemize}
\item[(i)]
			For both samples, calculate the MLE of the transition intensities $\hat\alpha^{(1)}$ and $\hat\alpha^{(2)}$ as given in \eqref{mle} and the corresponding test statistics  
			$\hat d^j:=| \hat\alpha^{(1)}_{0j} - \hat\alpha^{(2)}_{0j} |$, $j=1,\ldots,k$.
\item[(ii)]
{\bf Similarity test for state $j_0$}: 	
			For each state $j_0\in\{1,\ldots,k\}$ do:
		\begin{itemize}
		\item[(iia)]  In order to approximate the null distribution we define  constrained estimates $\overline \alpha^{(1)}, \overline \alpha^{(2)}$ of $\alpha^{(1)}$ and $\alpha^{(2)}$ minimizing the sum
		$\log \mathcal{L}_1 (\alpha^{(1)})+\log \mathcal{L}_2 (\alpha^{(2)})$
		of the log-likelihood functions defined 
		in \eqref{loglikelihood}   under the additional restriction 
			\begin{equation}\label{constr}
			d^{j_0}(\alpha^{(1)},\alpha^{(2)})= |\alpha^{(1)}_{0{j_0}} - \alpha^{(2)}_{0{j_0}}| = \Delta_{j_0},
			\end{equation}
			that is we estimate the transition intensities such that the models correspond to the margin of the (individual) null hypothesis \eqref{h2ind} for state $j_0$.
			Further define
				\begin{equation}\label{MLcons}
			\hat{\hat{\alpha}}^{(\ell)}_{0j}= \left\{
			\begin{array} {ccc}
			\hat \alpha^{(\ell)}_{0j} & \mbox{if} & \hat d^{j_0} \geq \Delta_{j_0} \\
			\overline \alpha^{(\ell)}_{0j} & \mbox{if} & \hat d^{j_0} < \Delta_{j_0}
			\end{array}  \right.,\ j=1,\ldots,k, \qquad \ell=1,2,
			\end{equation}
			and note that $\hat{\hat{\alpha}}^{(\ell)} = ( \hat{\hat{ \alpha}}^{(\ell)}_{01} , \ldots ,\hat{ \hat{ \alpha}}^{(\ell)}_{0k})^\top$.
			Consequently, if the test statistic $\hat d^{j_0}$ is larger or equal than the similarity threshold $\Delta_{j_0}$, which reflects the null situation, the original (and hence unconstrained) estimates $\hat\alpha^{(1)}$ and $\hat\alpha^{(2)}$ can be used.
			\item[(iib)] Use the constrained estimates 	$\hat{\hat{\alpha}}^{(\ell)},\   \ell=1,2$, derived in \eqref{MLcons}, to simulate bootstrap data $X^{*(1)}_1, \ldots, X^{*(1)}_{n_1}$
and  $X^{*(2)}_1, \ldots, X^{*(2)}_{n_2}$. Specifically we use the simulation approach as described in \citet{beyersmann2009}, where at first for all individuals survival times are simulated with all-cause hazard $\sum_{j=1}^{k}\hat{\hat{ \alpha}}^{(\ell)}_{0j}$ and then a multinomial experiment is run to decide on state $j$ with probability $\hat{\hat{ \alpha}}^{(\ell)}_{0j}/\sum_{j=1}^{k}\hat{\hat{ \alpha}}^{(\ell)}_{0j}$.
			\item[(iic)] For the datasets $X^{*(1)}_1, \ldots, X^{*(1)}_{n_1} $
and  $X^{*(2)}_1, \ldots, X^{*(2)}_{n_2}$ calculate
			the MLE $\hat\alpha^{*(1)}$ and $\hat\alpha^{*(2)}$ as in \eqref{mle} and the test statistic for state $j_0$ as in Step (i), that is
			$$\hat d^{*j_0}:=| \hat\alpha^{*(1)}_{0j_0} - \hat\alpha^{*(2)}_{0j_0}|.$$
								\end{itemize}
			\item[]	Repeat steps (iib) and (iic) $B$ times to generate $B$ replicates of the test statistic and let $\hat d^{*j_0(1)}, \ldots,  \hat d^{*j_0(B)}$ denote the corresponding order statistic.
			An estimate of the  $\alpha-$quantile of the distribution of 
			the statistic $\hat d^{*j_0}$ is then given by	$q_{\alpha}^*:=\hat d^{*j_0}_{(\lfloor B \alpha \rfloor )}$ and the null hypotheses in \eqref{h2ind} is rejected at the targeted significance level $\alpha$ whenever $\hat d^{j_0} < {q}_{\alpha}^{*}$.	 Alternatively a test decision can be made based on the $p$-value $\hat F^{j_0}_{B}(\hat d^{j_0}) = {\frac{1}{B}} \sum_{i=1}^{B}  I\{ \hat d^{*j_0(i)} \leq \hat d^{j_0} \}$, where $\hat F^{j_0}_{B}$ denotes the empirical distribution function of the bootstrap sample. 
						Finally we reject the individual null hypothesis \eqref{h2ind} for $j=j_0$ if $\hat F^{j_0}_{B}(\hat d^{j_0})<\alpha$ for a pre-specified significance level $\alpha$. 
\item[(iii)] The global null hypothesis in \eqref{h2} is rejected if 
	\begin{equation}\label{global_test}
	\max_{j_0=1,\ldots,k}{\hat F^{j_0}_{B}(\hat d^{j_0})}<\alpha.
	\end{equation}
\end{itemize}
\end{breakablealgorithm}

\bigskip

As stated above, the global null hypothesis \eqref{h2} is rejected if all individual null hypotheses are rejected. As a consequence of this procedure the power of the test decreases with an increasing size of states in the model as these are leading to a higher number of individual tests (see \citet{berger1982} for theoretical arguments on this). More precisely, it is a well known fact that methods based on the intersection union principle can be rather conservative (see for example \citet{phillips1990}), depending on the sample size, the variability of the data and the number of individual tests.
It can be shown that the test is consistent and controls its level. The theoretical arguments for that follow from adapting the proofs of \citet{detmolvolbre2015} to the present situation.
We will investigate the finite sample properties by means of a simulation study in Section \ref{sim}.

\subsection{Right-censoring}\label{censoring} 
Note that in case of right-censoring the methodology described above can be extended by adding corresponding factors from the distribution of the censoring times to the likelihood in \eqref{h9}. This requires the assumption of independence between censoring times and survival times. Under the assumption of independence the MLE in \eqref{mle} still remains valid. By estimating the censoring distribution from the data, Step (iib) in Algorithm \ref{alg1} can be conducted by additionally simulating (bootstrap) censoring times $C^{*(\ell)}_i$, $i=1,\ldots,n_\ell,\ \ell=1,2$, and defining the observed time as the minimum of the survival time and the censoring time.

\section{Simulation study}\label{sim}

\subsection{Design}

In the following we will investigate the finite sample properties of the proposed methods by means of a simulation study, driven by the application example given in Section \ref{appl}. We assume that individuals of two groups ($\ell=1,2$) are observed regarding three different outcomes over a period of $90$ days, hence we consider two competing risk models with each $j=3$ states over the time range $\mathcal{T}=[ 0,90 ]$. If there is no transition to one of the three states, an individual is administratively censored after these $90$ days.  The data in the following simulation study is generated according to the algorithm described in \citet{beyersmann2009}.

We consider in total four different scenarios, which are summarized in Table~\ref{tab_scen}. For the first three scenarios we choose the transition intensities of the application example in \eqref{estimates} (compare also Figure \ref{uromsm}). 
This choice results in true absolute differences of $$d^j=|\alpha^{(1)}_{0j}-\alpha^{(2)}_{0j}|=0.0002,0.0006,0.0005 \text{ for $j=1,2,3$},$$
which are also given in Table~\ref{tab_scen}.
In order to demonstrate the effect of different numbers of states, we start by testing for similarity of all three transition intensities simultaneously in the first scenario, whereas in the second and in the third scenario we only consider two states and hence only the difference of two transition intensities.
Precisely, in Scenario 2 we only compare the transition intensities for State 1 and 3 and in Scenario 3 we only consider State 1 and 2, respectively.
Finally, in the fourth scenario we choose identical models, that is $\alpha^{(1)}_{01}=\alpha^{(2)}_{01}=0.001,\ \alpha^{(1)}_{02}=\alpha^{(2)}_{02}=0.0011$ and $\alpha^{(1)}_{03}=\alpha^{(2)}_{03}=0.0004$, respectively, resulting in a difference of $0$ for all transition intensities. 

\begin{table}[]
\small
	\centering
	\begin{tabular}{c|c|c|c|c|c|c|c|c|c}
	\hline
	&	\multicolumn{3}{c}{Intensities model 1} & \multicolumn{3}{|c|}{Intensities model 2}  & \multicolumn{3}{c}{True absolute differences}\\
	\hline
         & $\alpha^{(1)}_{01}$ & $\alpha^{(1)}_{02}$ & $\alpha^{(1)}_{03}$ & $\alpha^{(2)}_{01}$ & $\alpha^{(2)}_{02}$ & $\alpha^{(2)}_{03}$ & $d^1$ & $d^2$ & $d^3$ \\
         	\hline
Scenario 1 & 0.001             & 0.0011            & 0.0004            & 0.0008   & 0.0017            & 0.0009 & 0.0002 & 0.0006 & 0.0005           \\
Scenario 2 & 0.001             & -                 & 0.0004            & 0.0008   & -                 & 0.0009  & 0.0002 & - & 0.0005          \\
Scenario 3 & 0.001             & 0.0011            & -                 & 0.0008   & 0.0017            & -     & 0.0002 & 0.0006 &   -         \\
Scenario 4 & 0.001             & 0.0011            & 0.0004            & 0.001    & 0.0011            & 0.0004   & 0 & 0 & 0       \\ 
		\hline
	\end{tabular}
		\caption{\small Transition intensities and their true absolute differences of the four different scenarios under consideration.} \label{tab_scen}
\end{table}

In other applications the number of patients ending up in one of the three states might be even smaller than the ones found in our application example. To this end, we consider a broader range of different sample sizes given by $$n=(n_1,n_2)=(200,200),(250,300),(300,300),(250,450),(300,500),(500,500),$$ where the choice of $(250,450)$ is the one closest to the application data in this paper and consequently the first three settings correspond to situations with less patients, particularly resulting in a smaller number of cases per state.
For example, choosing $n=(n_1,n_2)=(200,200)$ results for the first model after 90 days of observation in on average 16 patients in state 1, 18 patients in state 2 and 6 patients in state 3 and for the second model in 12 patients in state 1, 26 patients in state 2 and 14 patients in state 3, respectively (note that the numbers of patients have been rounded due to an easier interpretability). 

In order to simulate both the type I error and the power of the procedure described in Algorithm \ref{alg1}, we consider different similarity thresholds $\Delta=(\Delta_1,\Delta_2,\Delta_3)$ or, for scenarios 2 and 3, $\Delta=(\Delta_1,\Delta_2)$, respectively. Precisely we choose $\Delta_j\in\{0.00015,0.0002,0.0005,0.0006,0.001,0.0015,0.002\}$, where for the first three scenarios, the first four choices correspond to the null hypothesis \eqref{h2} and the other three to the alternative in \eqref{h3} (note that due to the sake of brevity not all choices are presented in the tables). Regarding the fourth scenario, we only consider $\Delta_j=0.001,\ 0.0015$ as in this case we only simulate the power of the test.

\subsection{Type I errors}\label{type1error}
Table \ref{tab1} displays the type I errors for scenarios 1-3. It turns out that the proportions of rejections of the null hypothesis \eqref{h2} for the global test are close to zero. These findings are in line with the theoretical arguments given after Algorithm \ref{alg1}, as tests based on the intersection union principle tend to be conservative in some situations. However it also becomes visible that this effect decreases when considering only two states instead of three (see the columns corresponding to Scenario 2 and 3, respectively).
Moreover we note that the individual tests yield a very precise approximation of the nominal level, as the proportion of rejections is close to $0.05$ in all scenarios under consideration. 

The difference between type I error rates of the individual tests and the global test become in particular visible when considering the first row of Figure~\ref{scenario1}, which yields a visualization of the results presented for Scenario 1 in Table~\ref{tab1}. Whereas the proportion of rejections are all around $0.05$ for the individual tests on all three states, the line indicating the results for the global test is close to zero. 

Finally, the points on the left of Figure~\ref{scenario1b}, corresponding to the smallest threshold, namely $\Delta=(0.00015,0.0002,0.0002)$, display the type I errors for a sample size of $n_1=n_2=300$ in a scenario which is not on the margin but in the interior of the null hypothesis. In this situation type I errors are smaller and well below the nominal level. Considering the individual test on the first state the proportion of rejection is close to $\alpha$ as the absolute distance $d^1=| \alpha^{(1)}_{01} - \alpha^{(2)}_{01} |=0.0002$, which is rather close to the chosen threshold $\Delta_1=0.00015$. For the other two states we have $d^2=0.0006$ and $d^3=0.0005$ and hence, regarding the similarity thresholds of $\Delta_2=\Delta_3=0.0002$, these situations correspond even stronger to the null situation, resulting in lower type I errors of the individual tests, given by $0.017$ for state $2$ and 0.009 for state $3$, respectively (compare Figure~\ref{scenario1b}).

\begin{table}[]
\small
	\centering
	\begin{tabular}{c|c|c|c}
	\hline
	\multirow{2}{*}{$(n_1,n_2)$ }	& Scenario 1 & Scenario 2 & Scenario 3\\
              & $\Delta=(0.0002,0.0006,0.0005)$ & $\Delta=(0.0002,0.0005)$  & $\Delta=(0.0002,0.0006)$  \\
		\hline
(200,200) & 0.000 (0.055/0.057/0.052) & 0.004 (0.047/0.048) & 0.004 (0.051/0.064) \\
(250,300) & 0.000 (0.066/0.049/0.048) & 0.002 (0.047/0.055) & 0.001 (0.049/0.057) \\
(300,300) & 0.000 (0.064/0.053/0.047) & 0.002 (0.048/0.058) & 0.002 (0.046/0.042) \\
(250,450) & 0.000 (0.051/0.058/0.063) & 0.004 (0.047/0.061) & 0.000 (0.038/0.051) \\
(300,500) & 0.001 (0.067/0.064/0.063) & 0.005 (0.038/0.062) & 0.004 (0.052/0.063) \\
(500,500) & 0.000 (0.053/0.052/0.062) & 0.005 (0.052/0.059) & 0.002 (0.041/0.062)\\
		\hline
	\end{tabular}
		\caption{\small Simulated type I errors of the test on similarity described in Algorithm \ref{alg1} for Scenarios 1-3 with $\Delta_j=d^j,\ j=1,2,3$, considering different sample sizes. The numbers in brackets correspond to the individual tests per state, the number outside to the global test result. The nominal level is chosen as $\alpha=0.05$.} \label{tab1}
\end{table}

\subsection{Power}
\begin{figure}[htb]
	\begin{center}
		\includegraphics[width=0.85\textwidth]{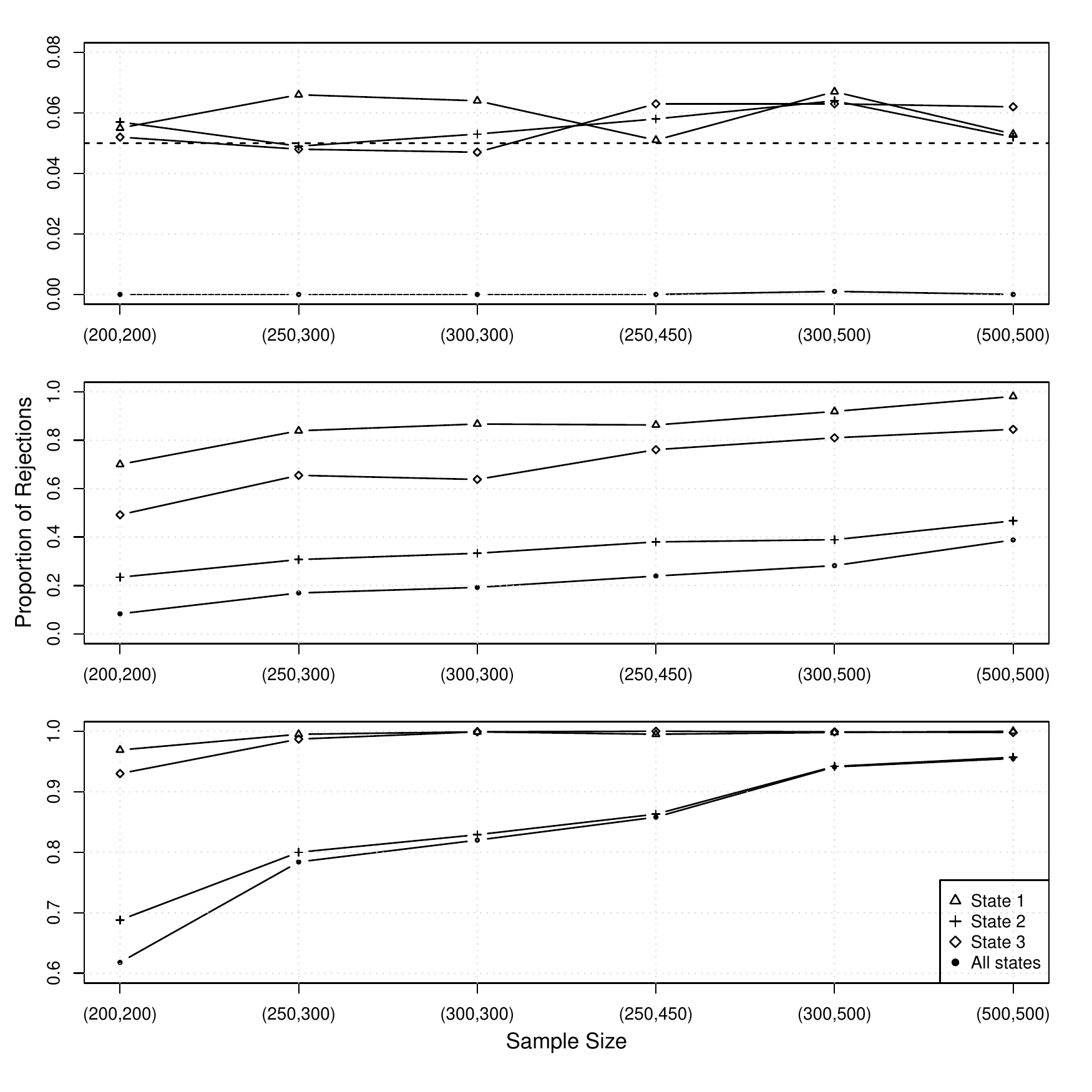}
		\caption{\small Proportion of rejections in dependence of the sample size for the individual tests on the three states and the global test, respectively, in Scenario 1. The three rows display different choices of $\Delta$, that is $\Delta=(0.0002,0.0006,0.0005)$ corresponding to the null hypothesis in the top row, $\Delta=(0.001,0.001,0.001)$ in the middle and $\Delta=(0.0015,0.0015,0.0015)$ in the bottom row, where the latter two correspond to the situation under alternative. The dashed line in the first row indicates the nominal level chosen as $\alpha=0.05$. }\label{scenario1}\end{center}
\end{figure}

Table \ref{tab_power} displays the simulated power of the global test and the individual tests for scenarios 1,2,3 and 4, respectively, as well as the two lower lines of Figure~\ref{scenario1} visualize some of the results from Scenario 1 of Table \ref{tab_power}. 
In general we observe that the test achieves a reasonable power in all scenarios under consideration and for increasing sample sizes the power converges to 1. For example, considering $n_1=n_2=300$ in Scenario 4, the simulated power lies between $0.837$ and $1.000$, depending on the threshold under consideration (see Scenario 4 in Table \ref{tab_power}). In particular keeping in mind the very small transition intensities (which result in only few cases in the several states, compare to the application example in Figure
~\ref{uromsm}) these results are very promising.
In addition, considering the first three scenarios it becomes obvious that the power increases significantly when just considering two instead of three states (compare for the same thresholds the results for Scenario 1 in Table~\ref{tab_power} to Scenarios 2 and 3). This becomes also visible in the two lower rows of Figure~\ref{scenario1} presenting the power of comparing states individually and simultaneously for the first scenario.
When assuming the same similarity thresholds $\Delta_1=\Delta_2=\Delta_3$ the power for the individual test on the second state is clearly below the observed values for the other two states. This results from the fact that the true absolute difference is given by $d^2=0.006$ and hence larger compared to $d^1$ and $d^3$, which are given by $0.002$ and $0.005$, respectively. 

When comparing the scenarios with only two states, that is Scenario 2 and Scenario 3 in Table \ref{tab_power}, we observe that the power of the global test is higher in the first. This holds for all sample sizes and choices of the threshold $\Delta$ and results from the different power obtained for the individual tests, which is due to the underlying assumed transition intensities. 
However, as mentioned beforehand, this effect decreases with increasing sample sizes, where, for all scenarios under consideration, the power converges to 1.

\begin{figure}[htb]
	\begin{center}
		\includegraphics[width=0.85\textwidth]{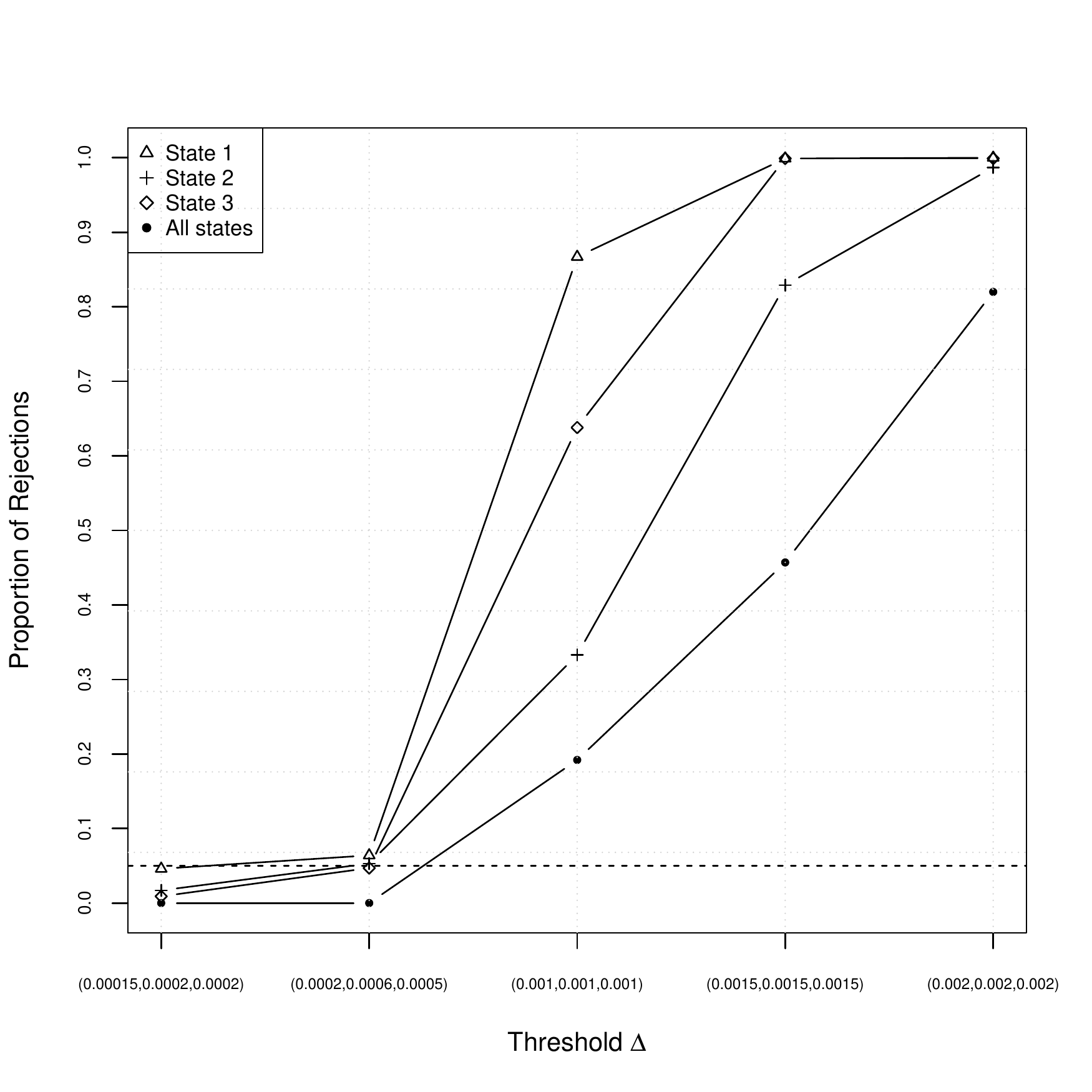}
		\caption{\small Proportion of rejections for a fixed sample size of $n_1=n_2=300$ in dependence of the threshold for the individual tests on the three states and the global test, respectively, in Scenario 1. The first two thresholds correspond to the null hypothesis (where the second one displays the margin situation), the last three to the alternative. The dashed line indicates the nominal level $\alpha=0.05$. }\label{scenario1b}\end{center}
\end{figure}

Finally, Figure~\ref{scenario1b} displays the proportion of rejections for Scenario 1 in dependence of the chosen similarity threshold $\Delta$. We observe that for the first two choices all values, that is the proportion of rejections for the individual test and the global test, respectively, are below or close to $\alpha$ as these situations correspond to the null hypothesis (see also the discussion at the end of Section \ref{type1error}).
For the other three choices of $\Delta$ presented in the right part of the figure simulations correspond to the alternative \eqref{h3}. Consequently, with increasing similarity thresholds, the proportion of rejections, which results in claiming similarity, increases.

\begin{table}[htb]
\small
	\centering
	\begin{tabular}{c|c|c|c}
	\hline
	\multicolumn{4}{c}{{\bf Scenario 1} } \\
	\hline
          $(n_1,n_2)$    & $\Delta=(0.001,0.001,0.001)$ & $\Delta=(0.001,0.0015,0.001)$  & $\Delta=(0.0015,0.0015,0.0015)$  \\
		\hline
(200,200) & 0.083 (0.700/0.234/0.492) & 0.217 (0.699/0.655/0.484) & 0.618 (0.969/0.688/0.930) \\
(250,300) & 0.169 (0.839/0.307/0.655) & 0.416 (0.836/0.787/0.622) & 0.784 (0.995/0.800/0.987) \\
(300,300) & 0.192 (0.867/0.333/0.638) & 0.457 (0.884/0.810/0.649) & 0.820 (0.999/0.829/0.999) \\
(250,450) & 0.239 (0.863/0.380/0.761) & 0.580 (0.852/0.893/0.753) & 0.858 (0.995/0.863/1.000) \\
(300,500) & 0.282 (0.919/0.389/0.810) & 0.701 (0.918/0.930/0.826) & 0.941 (1.000/0.942/0.999) \\
(500,500) & 0.388 (0.981/0.467/0.845) & 0.796 (0.982/0.948/0.851) & 0.955 (1.000/0.957/0.998)\\
		\hline
		\multicolumn{4}{c}{{\bf Scenario 2}} \\
		\hline
            & $\Delta=(0.001,0.001)$   & $\Delta=(0.001,0.0015)$ & $\Delta=(0.0015,0.0015)$   \\
		\hline
(200,200) & 0.382 (0.749/0.511) & 0.685 (0.723/0.952) & 0.932 (0.979/0.952) \\
(250,300) & 0.578 (0.864/0.669) & 0.852 (0.859/0.992) & 0.982 (0.994/0.988) \\
(300,300) & 0.594 (0.876/0.685) & 0.869 (0.876/0.991) & 0.989 (0.998/0.991) \\
(250,450) & 0.691 (0.887/0.782) & 0.883 (0.883/0.999) & 1.000 (1.000/1.000) \\
(300,500) & 0.765 (0.914/0.836) & 0.914 (0.914/1.000) & 1.000 (1.000/1.000) \\
(500,500) & 0.853 (0.984/0.866) & 0.984 (0.984/1.000) & 1.000 (1.000/1.000) \\
		\hline
		\multicolumn{4}{c}{{\bf Scenario 3}} \\
		\hline
            & $\Delta=(0.001,0.001)$   & $\Delta=(0.001,0.0015)$ & $\Delta=(0.0015,0.0015)$   \\
	\hline
(200,200)     & 0.184 (0.716/0.259) & 0.496 (0.716/0.683) & 0.666 (0.976/0.683) \\
(250,300)     & 0.266 (0.842/0.309) & 0.684 (0.840/0.813) & 0.803 (0.991/0.810) \\
(300,300)     & 0.264 (0.889/0.304) & 0.735 (0.890/0.830) & 0.827 (0.997/0.830) \\
(250,450)     & 0.318 (0.884/0.373) & 0.788 (0.884/0.900) & 0.899 (0.999/0.900) \\
(300,500)     & 0.374 (0.912/0.409) & 0.831 (0.912/0.915) & 0.913 (0.998/0.915) \\
(500,500)     & 0.432 (0.974/0.446) & 0.940 (0.975/0.965) & 0.968 (1.000/0.968) \\
			\hline
		\multicolumn{4}{c}{{\bf Scenario 4}} \\
		\hline
            & $\Delta=(0.001,0.001,0.001)$   & $\Delta=(0.001,0.0015,0.001)$ & $\Delta=(0.0015,0.0015,0.0015)$   \\
		\hline
(200,200) & 0.514 (0.757/0.685/0.981) & 0.738 (0.771/0.971/0.986) & 0.954 (0.986/0.968/1.000) \\
(250,300) & 0.760 (0.882/0.865/1.000) & 0.875 (0.881/0.994/0.991) & 0.996 (0.998/0.998/1.000) \\
(300,300) & 0.837 (0.936/0.896/0.998) & 0.930 (0.933/0.998/0.999) & 0.998 (1.000/0.998/1.000) \\
(250,450) & 0.856 (0.939/0.910/0.999) & 0.924 (0.927/0.998/0.998) & 1.000 (1.000/1.000/1.000) \\
(300,500) & 0.925 (0.968/0.956/1.000) & 0.953 (0.953/1.000/1.000) & 1.000 (1.000/1.000/1.000) \\
(500,500) & 0.982 (0.990/0.992/1.000) & 0.993 (0.993/1.000/1.000) & 1.000 (1.000/1.000/1.000) \\
		\hline
	\end{tabular}
		\caption{\small Simulated power of the test on similarity described in Algorithm \ref{alg1} for each scenario, considering different sample sizes and thresholds $\Delta$. The numbers in brackets correspond to the individual tests per state, the number outside to the global test result. The nominal level is chosen as $\alpha=0.05$.} \label{tab_power}
\end{table}

\clearpage

\section{Similarity of healthcare pathways involving prostate cancer surgery} \label{appl2}

We now want to address the question whether the readmission intensities for patients with prior in-house FB diagnostic are similar to the ones of the patients without prior in-house FB diagnostic (eq. \eqref{estimates}).
Therefore we perform the test on similarity described in Algorithm \ref{alg1} considering numerous different similarity thresholds $\Delta_j$, $j=1,2,3$, on the given application example. The choice of these thresholds is motivated from the simulation studies presented in Section \ref{sim}.
In Table \ref{tab6} we display the p-values of the individual tests on states $1$, $2$ and $3$, respectively, for six different similarity thresholds. 
We observe that for the smallest threshold, that is $\Delta_j=0.0005$, all individual p-values are far above the nominal level of $\alpha=0.05$. 
For $\Delta_j=0.0007$ the individual p-value of the test for the first state is now given by $0.044$, which is below the nominal level and results in claiming similarity of transition intensities for the first state. Considering the same threshold for state 2 and 3, respectively, yields that for these states the individual null hypotheses cannot be rejected.
Further, considering $\Delta_j=0.0008$ we observe that now similarity of the corresponding readmission intensities can be claimed for state 1 and 3, as both individual p-values are below the nominal level. The same holds for $\Delta_j=0.001$, as the p-values for state 1 and 3 are given by $0.006$ and $0.004$, respectively, whereas the p-value of the test for state 2 is given by $0.094$. 
However, since for both thresholds, that is $\Delta_j=0.0008$ and $\Delta_j=0.001$, each p-value for state 2 is larger than $\alpha=0.05$, the global null hypothesis in \eqref{h2} cannot be rejected according to the decision rule \eqref{global_test}.
For the two largest choices of $\Delta_j$ given by $0.0012$ and $0.0015$ respectively, all individual p-values are well below $\alpha=0.05$ which means that the global null hypothesis \eqref{h2} can be rejected and similarity can be claimed for all three states, that is we decide for similarity of both patient populations regarding all their readmission intensities.
Finally we observe that the same conclusion can be made for all thresholds $\Delta$ fulfilling $\Delta_1\geq 0.0007$, $\Delta_2\geq 0.0012$ and $\Delta_3\geq 0.0008$ as this choice guarantees that all individual p-values are below the nominal level of $\alpha=0.05$.
In terms of difference in number of events this translates as follows: Assuming for example two samples of 350 patients each and follow-up of 90 days, these thresholds correspond to allowing for a difference of approximately at most 22, 38, and 25 events for transitions into states 1, 2, and 3 between both groups. 

\begin{table}[tb]
\begin{tabular}{c|c|c|c|c|c|c}
\hline    
 \multirow{2}{*}{State} & \multicolumn{6}{c}{Similarity threshold $\Delta_j$}             \\ 
            & 0.0005 & 0.0007 & 0.0008 & 0.0010 & 0.0012 & 0.0015  \\\hline    
State 1     &   0.166    & \bf{0.044} &    \bf{0.026}     &    \bf{0.006} &   \bf{0.002}    &  \bf{$<$0.0001}             \\
State 2     &    0.514  & 0.366        &   0.251       &     0.094   &   \bf{0.037}    &     \bf{$<$0.0001}           \\
State 3     &   0.502    & 0.104       &   \bf{0.045}     & \bf{0.004} &   \bf{$<$0.0001}      &   \bf{$<$0.0001}          \\\hline  
\end{tabular}
	\caption{\small P-values of the individual tests on similarity described in Algorithm \ref{alg1} for the application example considering different thresholds $\Delta_j$. Bold values indicate p-values below the nominal level of $\alpha=0.05$.} \label{tab6}
\end{table}

\section{Discussion} \label{disc}

In this paper we developed a hypothesis test based on a
constrained (parametric) bootstrap to assess the similarity of competing risk models with constant transition intensities. Specifically, we performed an individual test for each state and combined these individual tests by applying the intersection union principle. We examined the finite sample properties by numerous simulations motivated by an example application in urology, and demonstrated that the test properly controls its level and yields a reasonable power. It would be interesting to investigate further whether the power can be improved even more by not performing $k$ individual tests, but by defining a global test statistic that directly accounts for all states. This alternative test statistic might yield a procedure with increased power but comes at the cost of not being able to draw conclusions for each state individually as all information from the different states is summarized in one quantity.  

We proposed measuring similarity by the absolute difference between transition intensities. However, instead of considering  differences a similar methodology can be developed for comparing the ratios of the transition intensities. In the case of the application example, this would mean examining the following ratios to test for similarity: $\alpha_{01}^{(1)}/\alpha_{01}^{(2)} = 1.25,\ \alpha_{02}^{(1)}/\alpha_{02}^{(2)} = 0.65,$ and $ \alpha_{03}^{(1)}/\alpha_{03}^{(2)} = 0.44$. On the one hand, considering ratios would have the advantage that they are time-invariant, i.e., not depending on the time scale anymore. On the other hand, for small transition intensities, i.e., settings with few events, differences may better communicate situations where there is no large difference in terms of intensities or events as compared to ratios. For instance, while $|\alpha_{03}^{(1)} - \alpha_{03}^{(2)}| = 0.0005$ is in fact fairly small this may by far not be assumed when examining the ratio $\alpha_{03}^{(1)}/\alpha_{03}^{(2)} = 0.44$. In general, the choice how to measure the deviation between the transition intensities depends on the goal of the study and should be carefully investigated by the researcher. This also applies to the corresponding equivalence thresholds which offer on the one hand a maximum of flexibility for our approach but on the other hand also provide the need of a very careful discussion in advance. Currently there are no guidelines fixing these thresholds in studies as considered in Section \ref{appl}, which makes this decision an important topic for further research.

With respect to the application example, we were able to identify thresholds for which the global null hypothesis could be rejected and therefore the transition intensities are to be considered similar. The chosen thresholds were based on the results of the simulation study and were also chosen differently to illustrate the effect on the p-values. This extensive procedure is not necessary for future applications of the method to clinical data, instead a careful preliminary determination of plausible equivalence thresholds is required. This can be done in such a way that one considers which difference in number of events one would still like to allow and then calculates the corresponding threshold accordingly, taking into account the examined time span and sample size. We point out here that while very stringent thresholds are often expected from an equivalence test of a therapeutic study, these would rather not be fulfilled in our application example. However, we can consider this somewhat less stringent and continue to assume similarity, since the actual goal is to use the overall data to examine an outcome that is supposed to be no longer directly related to the diagnostic procedure. A further strength of the kind of data we used in our application is that it was drawn from sets of claims data having a standardized format used for quality assurance and for the calculation of the German Diagnosis Related Groups system. The process and quality assurance measures for providing this dataset are highly standardized. The data are easily accessible and therefore provide a good source of information for this investigation.

A limitation of the methodology proposed in this article is the assumption of constant transition intensities, which may not be met in real data applications. 
However, our proposed approach based on parametric bootstrap allows in principle an extension to different parametric distributions of event times. This requires further extensive investigations, which are beyond the scope of this work. We therefore leave it for future research.


\bibliography{ref.bib}
\bibliographystyle{statinmed}

\end{document}